# Nanocrystal Growth on Graphene with Various Degrees of Oxidation

Hailiang Wang, Joshua Tucker Robinson, Georgi Diankov, and Hongjie Dai[*]
*Department of Chemistry and Laboratory for Advanced Materials, Stanford University, Stanford, CA 94305, USA*
E-mail: hdai@stanford.edu

As a single-atom thick carbon material with light-weight, high surface area and conductivity, graphene[1,2] could be ideal substrates for growing and anchoring of functional nanomaterials for high performance electrocatalytic or electrochemical devices. Nanocrystals grown on graphene could have enhanced electron transport rate, high electrolyte contact area and structural stability, all of which could be useful for various fundamental and practical applications.[3] Although decoration of nanoparticles on graphite oxide (GO) sheets has been shown,[4-6] it remains unexplored and highly desirable to synthesize nanocrystals on more pristine graphene with high electrical conductivity, control the morphologies of the nanocrystals by tuning the oxidation degrees of the graphene sheets, and rationalize the nanocrystal growth behavior.

Here, we show a general two-step method to grow hydroxide and oxide nanocrystals of the iron family elements (Ni, Co, Fe) on graphene with two degrees of oxidation. Drastically different nanocrystal growth behaviors were observed on low-oxidation graphene sheets (GS) and highly oxidized GO in hydrothermal reactions. Small particles pre-coated on GS with few oxygen-containing surface groups diffused and recrystallized into single-crystalline nanoplates or nanorods with well defined shapes. In contrast, particles pre-coated on GO were pinned by the high-concentration oxygen groups and defects on GO without recrystallization into well-defined morphologies. Our results suggest an interesting approach to controlling the morphology of nanocrystals by tuning the surface chemistry of graphene substrates used for crystal nucleation and growth.

Our GS with low degree of oxidation were made by an exfoliation-reintercalation-expansion method,[7-9] and GO was produced by a modified Hummers method[10] (Figure 1). The resistivity of our

GS was measured to be only several times higher than pristine graphene, but ~100 times lower than GO in reduced form.[7-9] The oxygen content in GS (~5%) was much lower than in GO (~20%) as measured by Auger spectroscopy[9] and X-ray photoelectron spectroscopy (XPS).[11]

In the first step of $Ni(OH)_2$ growth on graphene, we deposited precursor materials in the form of small nanoparticles uniformly nucleated onto GS or GO by hydrolysis of $Ni(CH_3COO)_2$ at 80℃ in a 10:1 *N,N*- dimethylformamide (DMF)/water mixture (see Supporting Information). The 10:1 DMF/$H_2O$ ratio was found important to afford good dispersion of graphene and slow rate of hydrolysis that led to selective and uniform coating of nickel hydroxide on graphene, with little particle growth in free solution. After the first step, dense and uniform $Ni(OH)_2·0.75H_2O$ nanoparticles (~10-20nm in diameter) were formed on both GS and GO, revealed by scanning electron microscopy (SEM) (Figure 2a, 3a) and X-ray diffraction (XRD) (Figure S1a,1b). The percentage of mass of $Ni(OH)_2·0.75H_2O$ was ~87% in the graphene composite. As a control, when water was used as the sole solvent, we observed appreciable $Ni(OH)_2·0.75H_2O$ particle growth in free solution instead of on GS due to fast hydrolysis.

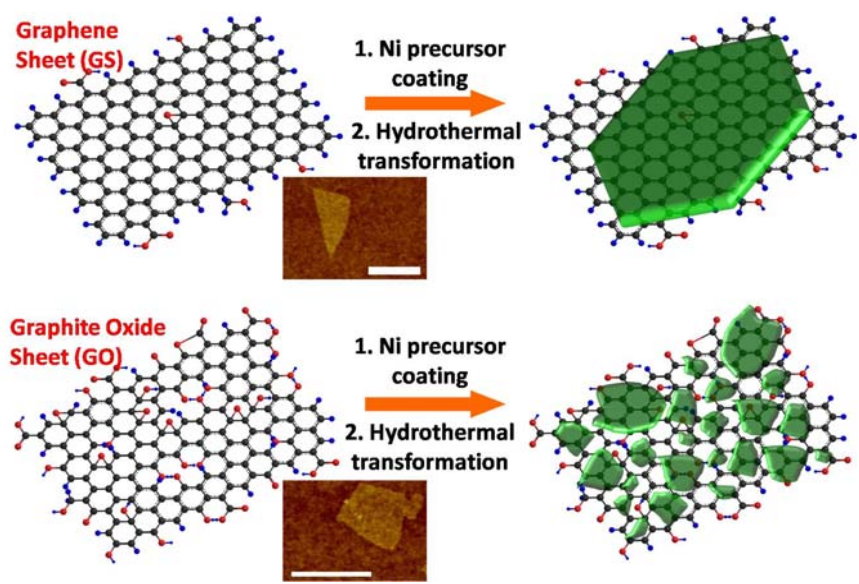

**Figure 1.** Schematic two-step $Ni(OH)_2$ nanocrystal growth on graphene sheets (GS, top panel) and graphite oxide (GO, bottom panel). Dark grey balls: carbon atoms; blue balls: hydrogen atoms; red balls: oxygen atoms; green plates: $Ni(OH)_2$. Insets: AFM images of GS and GO. Scale bars: 500nm. After the first step of reaction (Ni precursor coating), the same coating of $Ni(OH)_2·0.75H_2O$ was obtained both on GS and GO. After the second step of reaction (hydrothermal transformation), however, the coating on GS diffused and recrystallized into large single-crystalline hexagonal $Ni(OH)_2$ nanoplates, while the coating on GO remained as densely packed nanoparticles pinned by the functional groups and defects on GO surface.

In the second step, we hydrothermally treated the first step product, i.e. Ni(OH)$_2$·0.75H$_2$O coated GS at 180℃ in water (see Supporting Information). We observed that the coating material evolved from dense small particles into hexagonal nanoplates selectively attached to GS (Figure 1 top panel and Figure 2b,2c). The side length of the nanoplates was several hundred nanometers with thickness <~10 nm (Figure 2b,2c, Figure S2). XRD of a thick layer of packed nanoplates/GS suggested crystalline β-Ni(OH)$_2$ formed on graphene (Figure 2d). High resolution transmission electron microscopy (HRTEM) (Figure S3) clearly revealed the (100) and (010) lattice fringes in the plane of single-crystalline hexagonal Ni(OH)$_2$ nanoplate on GS. The corresponding fast Fourier transform (Figure S3 inset) of the high resolution TEM image was consistent with hexagonal lattice perpendicular to the (001) zone axis, suggesting Ni(OH)$_2$ nanoplates attached to GS at their (001) planes. In a film of the packed Ni(OH)$_2$/GS plates for XRD experiments, a large fraction of the plates were packed in parallel to each other and to the substrate, giving an enhanced (001) diffraction peak in the XRD spectrum (Figure 2d). Scanning Auger electron spectroscopy (SAES) elemental imaging of Ni and C elements in Ni(OH)$_2$/GS composite further confirmed attachment of Ni(OH)$_2$ nanoplates on GS (Figure S4).

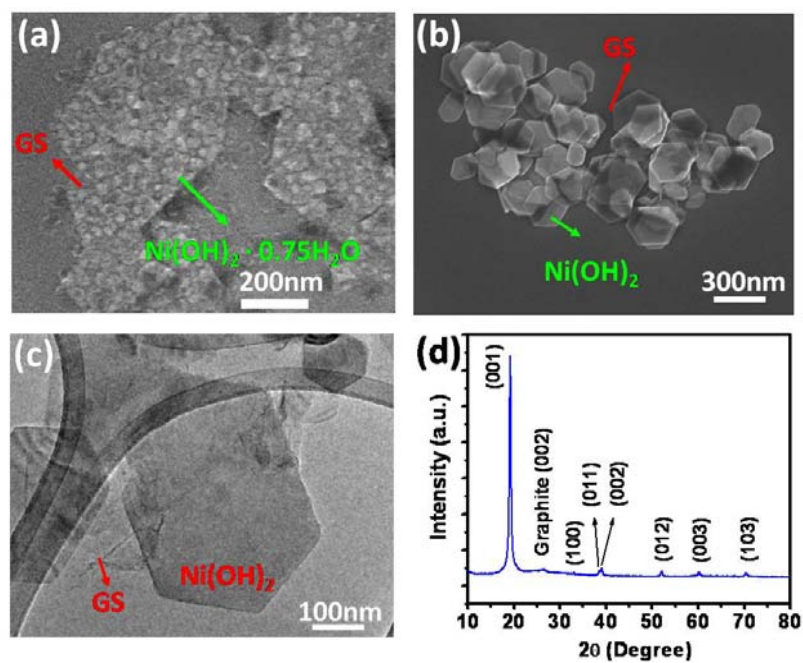

**Figure 2.** (a) An SEM image of Ni(OH)$_2$·0.75H$_2$O particles uniformly coated on GS after the first step of reaction at 80℃. (b) An SEM image of Ni(OH)$_2$/GS after a second step of simple hydrothermal treatment of the product depicted in (a) at 180℃. (c) A TEM image of hexagonal Ni(OH)$_2$ nanoplates formed on top of GS. (d) XRD spectrum of a packed thick film of hexagonal Ni(OH)$_2$ nanoplates on GS.

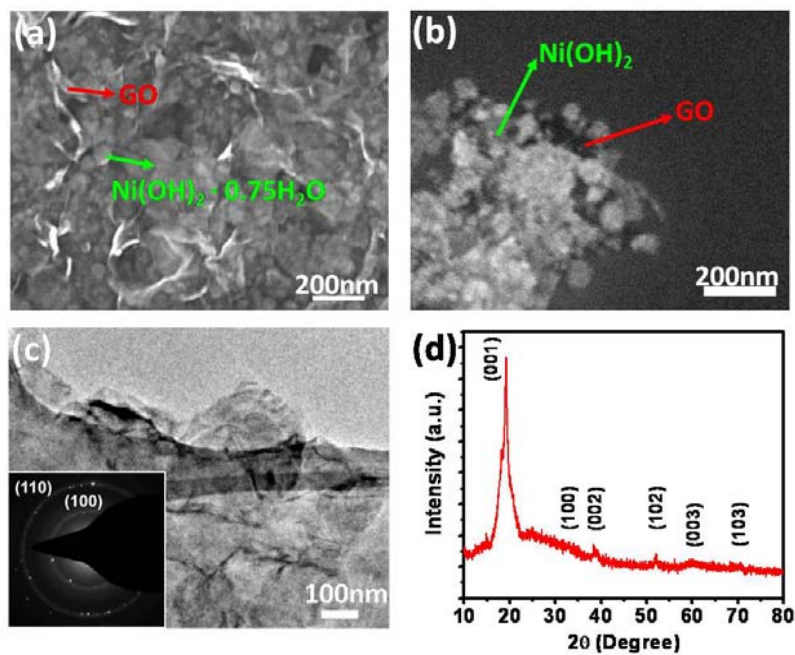

**Figure 3.** (a) An SEM image of Ni(OH)$_2$·0.75H$_2$O/GO after the first step of reaction at 80℃. (a) An SEM image of Ni(OH)$_2$/GO after the second step of reaction at 180℃. (c) A TEM image of Ni(OH)$_2$/GO span over the holes on a TEM grid, inset of (c), an electron diffraction pattern of Ni(OH)$_2$ on GO. (d) XRD spectrum of a packed thick film of Ni(OH)$_2$ nanoparticles on GO.

The same Ni(OH)$_2$·0.75H$_2$O coating obtained on GO transformed into small nanoparticles of β-Ni(OH)$_2$ densed packed on GO (forming a continuous film) after the second step of hydrothermal treatment at 180℃ (Figure 1 lower panel, Figure 3), without producing large single-crystalline hexagonal nanoplates as in the GS case. These results showed that the size, morphology and crystallinity of nanocrystals formed on graphene were dependent on the degrees of oxidation of the underlying graphene substrates. We suggest that GS with fewer functional groups and defects exhibit weaker chemical interactions with coating species on the surface. During the 180℃ hydrothermal reaction, the small coating particles on GS diffused across the 'slippery' graphitic lattice and recrystallized into single- crystalline hexagonal nanoplates on the GS. On GO, however, due to higher density of oxygen functional groups including carboxylic, hydroxyl, and epoxy groups,[7,9-12] the GO surface interacted strongly with the coated species, providing pinning forces to the small particles to hinder diffusion and recrystallization. As a result, most of the Ni(OH)$_2$·0.75H$_2$O particles coated on GO by the first step reaction remained pinned at the original positions after the second step hydrothermal treatment at higher temperature. In the case of Ni(OH)$_2$/GS, it is plausible that both chemisorption and van der Waals interactions coexist between Ni(OH)$_2$ nanoplates and graphene, at oxygen-containing defect sites and pristine regions of the GS respectively.

We found it a general approach to control nanocrystal morphology by two step synthesis on graphene with different degrees of oxidation (Figure S5-S8). By controlling the second step reaction temperature, we produced CoO(OH) and Fe$_2$O$_3$ nanocrystals with regular nanoplate and nanorod morphologies on GS, using Co(CH$_3$COO)$_2$ and Fe(CH$_3$COO)$_2$ as precursors (Figure S5c, Figure S7c) respectively. On GO, only small irregularly shaped nanoparticles were resulted (Figure S5f, Figure S7f). These results confirmed graphene with various degrees of oxidation can be used as a novel substrate for the growth of nanocrystals into various sizes and morphologies. Adjusting the second step reaction temperature can be combined to further control materials grown on graphene. For particles with weak interactions with graphene, increasing the reaction temperature eventually led to diffusion and recrystallization of surface species into larger crystals on GO (Figure S9).

In summary, we developed a two-step method to grow nanocrystals with well-defined nanoplate or nanorod morphologies on weakly interacting and highly conducting graphene surfaces. The morphology of the nanocrystals formed on graphene can be tailored by the oxidation degree of graphene and reaction temperature, which could be extended to synthesize a wide range of functional nanomaterials.